\documentclass[%
 aip,
 amsmath,amssymb,
 reprint,%
]{revtex4-1}

\usepackage{graphicx,xcolor}
\usepackage{dcolumn}
\usepackage{bm}
\usepackage[percent]{overpic}

\usepackage[utf8]{inputenc}
\usepackage[T1]{fontenc}
\usepackage{mathptmx}
\usepackage{etoolbox}

\makeatletter
\def\@email#1#2{%
 \endgroup
 \patchcmd{\titleblock@produce}
  {\frontmatter@RRAPformat}
  {\frontmatter@RRAPformat{\produce@RRAP{*#1\href{mailto:#2}{#2}}}\frontmatter@RRAPformat}
  {}{}
}%
\makeatother
\begin{document}


\title{Percolation in binary mixtures of  linkers and particles: chaining {\it {vs}} branching}
\author{M. Gouveia}
 \affiliation{Centro de F\'{\i}sica Teórica e Computacional, Universidade de Lisboa,
1749-016 Lisboa, Portugal.}
 \affiliation{Departamento de F\'{\i}sica , Faculdade de Ci\^encias, Universidade de Lisboa,
1749-016 Lisboa, Portugal.}
 \email{mgouveia@alunos.fc.ul.pt}
\author{C.S. Dias}%
 \email{csdias@fc.ul.pt}
\affiliation{Centro de F\'{\i}sica Teórica e Computacional, Universidade de Lisboa,
1749-016 Lisboa, Portugal.} 
\affiliation{Departamento de F\'{\i}sica , Faculdade de Ci\^encias, Universidade de Lisboa,
1749-016 Lisboa, Portugal.}

%

\author{J.M. Tavares}
\affiliation{Centro de F\'{\i}sica Teórica e Computacional, Universidade de Lisboa,
1749-016 Lisboa, Portugal.}%
\affiliation{Instituto Superior de Engenharia de Lisboa, ISEL,
Avenida Conselheiro Em\'{\i}dio Navarro, 1 1950-062 Lisboa, Portugal}
\email{jmtavares@fc.ul.pt}
\date{\today}

\begin{abstract}
Equilibrium gels of colloidal particles can be realized through the introduction of a second species, a linker that mediates the bonds between the colloids.
A gel forming binary mixture whose linkers can  self-assemble into linear chains while still promoting the aggregation of particles is considered in this work. The particles are patchy particles 
with $f_C$ patches of type $C$ and the linkers are patchy particles 
with $2$ patches of type $A$ and $f_B$ patches of type B. 
The bonds between patches of type $A$ ($AA$ bonds) promote the formation of linear chains of linkers. 
Two different ways (model A and model B) of bonding the linkers to the particles - or inducing branching - are studied. In model A, there is a competition between chaining and branching, since the bonding between linkers and particles is done through $AC$  bonds only. In model B linkers aggregate to particles through bonds $BC$ only, making chaining and branching independent.  
The percolation behaviour of these two models is studied in detail, employing a generalized Flory-Stockmayer theory and Monte Carlo simulations. 
The self-assembly of linkers into chains 
reduces the fraction of particles needed for percolation to occur (models A and B) and induces percolation when the fraction of particles is high (model B). Percolation by heating and percolation loops in temperature composition diagrams are obtained when the formation of chains is energetically favourable, by increasing the entropic gain of branching (model A).  
Chaining and branching are found to follow a model dependent relation at percolation, which shows that, for the same composition, longer chains require less branching for percolation to occur.

\end{abstract}

\maketitle

\section{\label{sec:intro}Introduction}


The theoretical and simulation studies of patchy particle models \cite{bianchi2006,sciortino2011,tavares2009,teixeira2017} have opened the way to the concept of equilibrium gels, a thermodynamic stable phase formed by a percolated network of particles, that is not related to phase separation arrest. These works have shown that the formation of  bonds between particles with low functionality leads to the emergence of a percolated fluid at low temperatures and low densities, that does not phase separate with a vapor.
Experimental evidence of the existence of equilibrium gels was first found in single component systems like laponite \cite{ruzicka2011} (a colloidal clay), DNA nanostars\cite{biffi2013}  and a solution of
Fmoc-di\-phe\-ny\-la\-la\-ni\-ne molecules in dimethyl sulfoxide\cite{dudukovic2014}.  
 
In recent years, much attention has been devoted to the study of aggregation in binary mixtures where the self-assembly of one of the components is mediated and controlled by the other. Examples span a large variety of soft matter and biological systems, like, amongst many others, protein-protein aggregation \cite{heidenreich2020},  cross linking of actin filaments \cite{Muller2014}, colloidal dispersions and protein fibrils \cite{Peng2016}, nanoparticles and globular proteins \cite{Bharti2014}, microparticles and small soft microgels \cite{Luo2015}, and cell-mediated colloidal scaffolds \cite{Dias2020,Custodio2015,Custodio2014}. 
Following a similar line of reasoning, 
the search for realizations of equilibrium gels with an extra degree of control has led to the study of model systems where  bonds between particles are mediated by another component, the linkers \cite{lindquist2016,howard2019,xia2020,lowensohn2019,teixeira2021}. In these works, the linkers are bifunctional (i.e. they bond two particles) and can be polymers of different lengths \cite{howard2019}, DNA strands \cite{xia2020,lowensohn2019}, or patchy particles \cite{lindquist2016,teixeira2021}. In any case, their properties (size, shape, and concentration) can be used to control  the aggregation process and the thermodynamics, and can be adjusted to obtain the proper conditions for the formation of equilibrium gels. These works  establish some interesting results for linker-particle aggregating systems: a percolated network appears at low temperatures in a finite range of linker concentrations \cite{lindquist2016,teixeira2021} that depends on the functionality of particles; the connectivity properties of this network can be controlled by the amount of linkers \cite{lindquist2016,howard2019,teixeira2021}; it is possible to find single phase percolated fluids at low densities and temperatures \cite{lindquist2016,howard2019,teixeira2021}, and these densities can be further reduced using longer linkers \cite{howard2019}. An experimental system\cite{song2020} formed by nano-particles and  telechelic polymer chains (the linkers) confirms some of these predictions, since it is found that the gelation of particles is controlled by the relative concentration of polymeric linkers. 

The aim of the present work is to thoroughly investigate the percolation thresholds of  linker-particle patchy models  in which  linkers can  also assemble into linear chains, using theory and simulation. This is accomplished by introducing bonds between linkers. While linker-linker bonds  promote chaining, the usual particle-linker bonds lead to the branching of those chains and eventually to the formation of percolated networks. Two types of interplay between chaining and branching are addressed (in two variants of this model). (i) a competition between chaining and branching is set by letting the two patches of the linkers bond either to another linker or to a particle (model A); (ii) chaining and branching are set independently,  by letting the linkers have two patches that bond only to linkers and other patches that bond only to particles (model B). It will be shown   that 
the chaining of linkers affects strongly the conditions at which percolation occurs, and that these changes are different in models A and B. 
It should be emphasized that percolation is only a necessary condition to obtain equilibrium gels. The investigation of the phase diagrams of the percolated phases of these models is left for future work. Still, it is important to stress that in almost all models of binary mixtures of patchy particles with low valence 
\cite{lindquist2016,teixeira2021,delasHeras2011a,delasHeras2011b,delasHeras2012}, the phase diagrams (calculated using Wertheim's theory) always exhibit  percolated  single phases  at low temperatures for a range of low densities. 

The paper is organized as follows. In section II  the patchy particle models are introduced in detail, and the simulation and theoretical methods employed to study the percolation properties of the models are described. In section III  the percolation thresholds obtained for the two models are described and analysed. Finally, in section IV the results are discussed and conclusions are drawn.

\section{\label{sec:model}Model}
\begin{figure}[htb]
\begin{overpic}[width=0.2\textwidth]{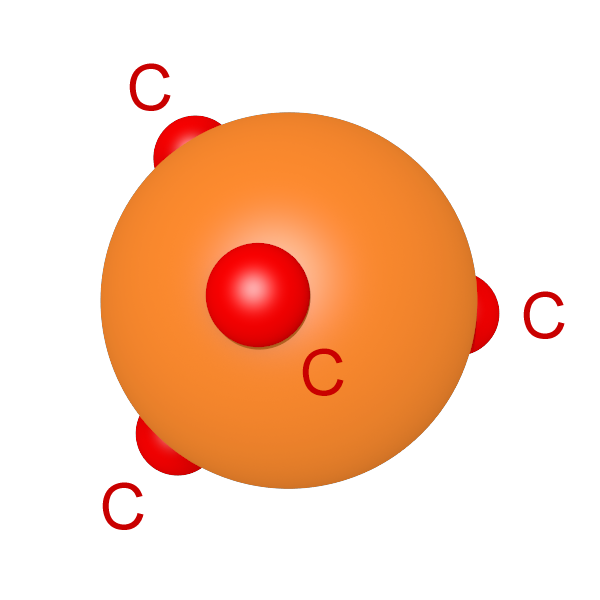}
 \put (2,90) {a)}
\end{overpic}
\begin{overpic}[width=0.2\textwidth]{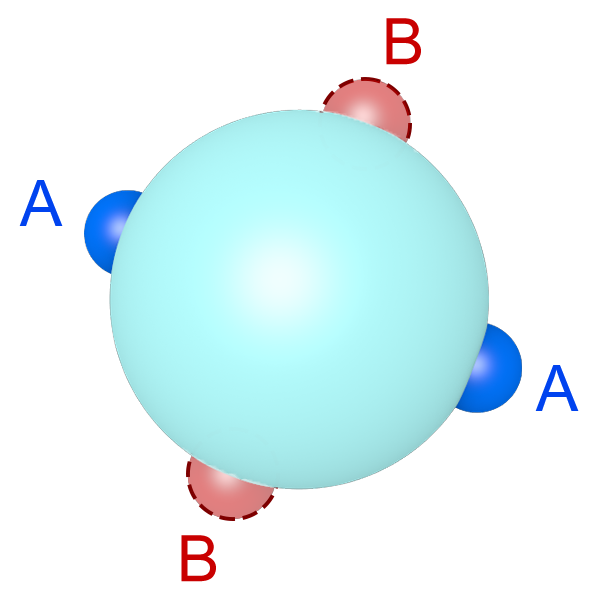}
 \put (2,90) {b)}
\end{overpic}
\caption{Species of the binary mixture. a) Particle: hard sphere with diameter $\sigma$ and $f_C$ patches of type C on its surface.  b) Linker: hard sphere with diameter $\sigma$ and 2 patches of type A and $f_B$ patches of type B on its surface.}
\label{fig:particles}
\end{figure}

A binary mixture of $N_P$ particles and $N_L$ linkers in a volume $V$ is considered. Particles and linkers are hard spheres (HSs) of diameter $\sigma$. The (reduced) density of the system is $\rho = (N_L+N_P)\sigma^3/V$ and the composition $x$ is the fraction of particles, $x=N_P/(N_L+N_P)$. Particles  are decorated with $f_C\ge 3$ patches of type C on its surface,
while linkers are decorated with 2 patches of type A 
and up to $f_B$ patches of type B  (see Fig.~\ref{fig:particles}). 

The interaction potential $V_{ij}$ between HSs $i$ and $j$ is,
\begin{equation}
\label{potential}
V_{i j}=V_{H S}\left({r}_{i j}\right)+\sum_{(\alpha, \beta) \in \Gamma_{i j}} V_{\alpha \beta}({r}_{i j}^{\alpha \beta}),
\end{equation}
where $V_{HS}$ is the HS potential,
$V_{\alpha\beta}$ is a spherical-well potential (of energy depth $-\epsilon_{\alpha\beta}\le0$ and range $\delta_{\alpha\beta}$) corresponding to the interaction between a patch of type $\alpha$ on HS $i$ and a patch of type $\beta$ on HS $j$ from the set $\Gamma_{ij}$ of possible such pairs,  ${r}_{ij}$ is the distance between HSs $i$ and $j$, and ${r}_{ij}^{\alpha\beta}$ is the distance between the  center of patch $\alpha$ in HS $i$ and the center of patch $\beta$ in HS $j$ \cite{bianchi2008}. Essentially, a bond $\alpha \beta$ between HSs $i$ and $j$ is established and the potential energy decreases by $\epsilon_{\alpha\beta}$ when ${r}_{ij}^{\alpha\beta}<\delta_{\alpha\beta}$.
In simulations, the placement of the patches over the HS and the ranges $\delta_{\alpha\beta}$ of the patch-patch potential are chosen so that it is guaranteed that each patch engages at most in a single bond \cite{bianchi2008}. 
\begin{figure}[htb]
\begin{overpic}[width=0.2\textwidth]{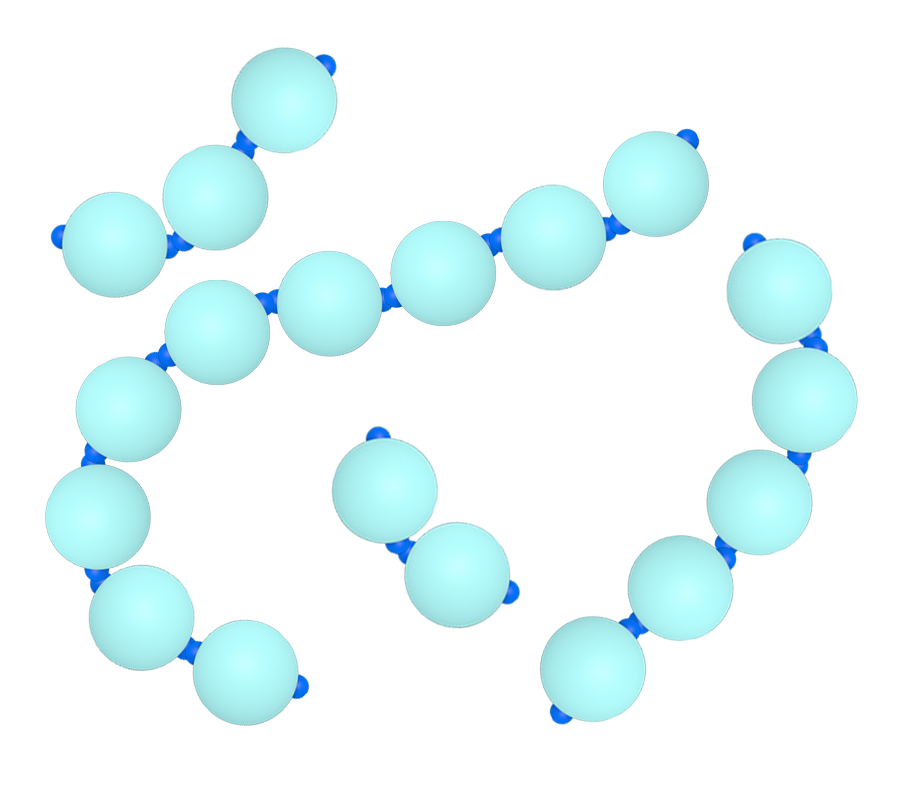}
 \put (5,80) {a)}
\end{overpic}
\begin{overpic}[width=0.2\textwidth]{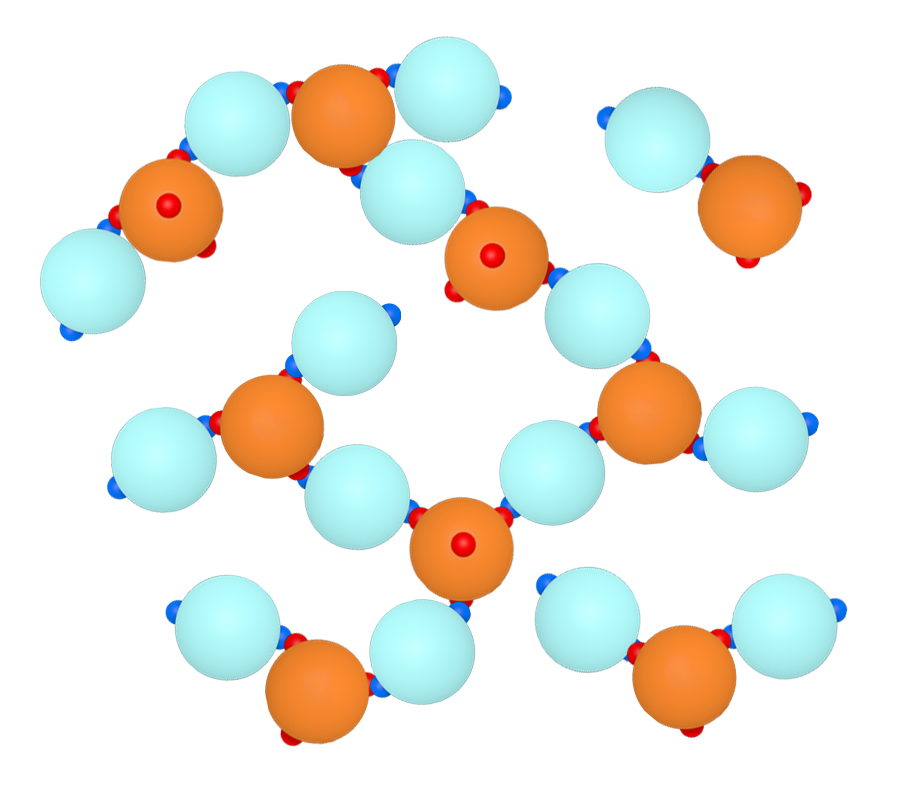}
 \put (5,80) {b)}
\end{overpic}
\begin{overpic}[width=0.2\textwidth]{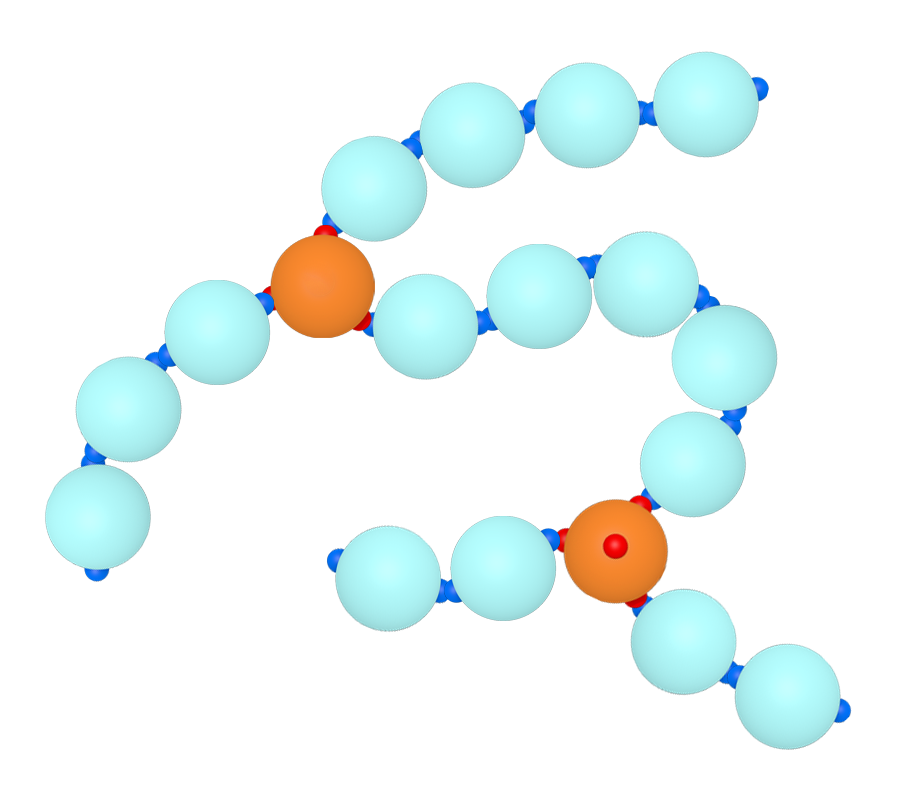}
 \put (5,80) {c)}
\end{overpic}
\begin{overpic}[width=0.2\textwidth]{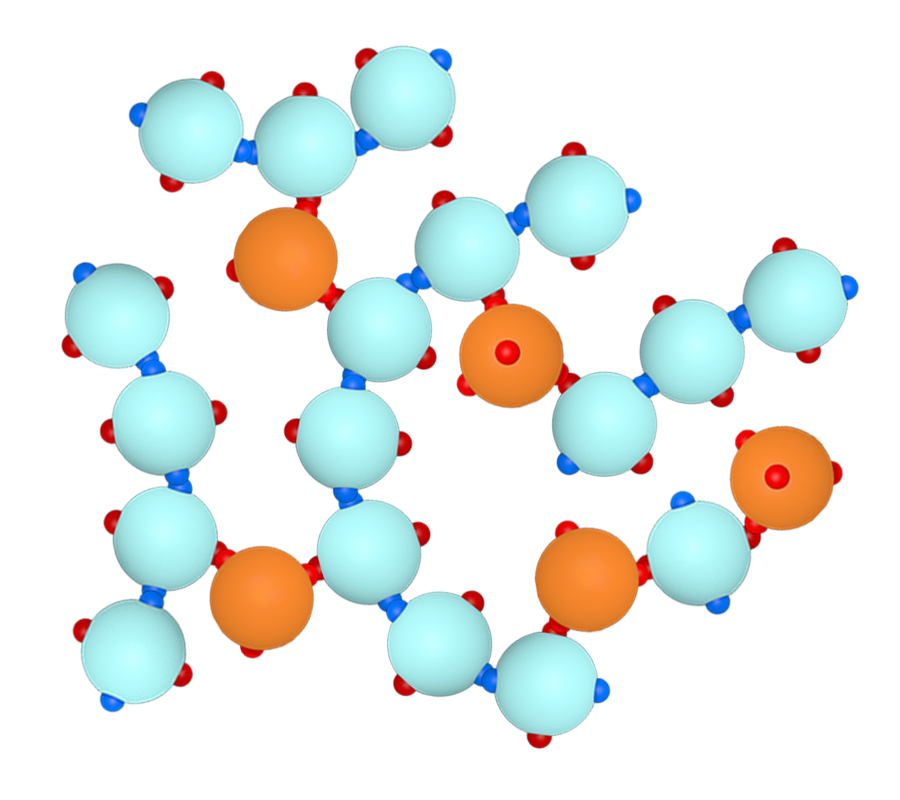}
 \put (5,80) {d)}
\end{overpic}
\caption{Typical clusters obtained in different models of mixtures constituted by the species depicted in Fig.~\ref{fig:particles}. a): chains of linkers (only $\epsilon_{AA} \ne 0$). b): 
clusters where  a single linker mediates the bonds between the particles \cite{lindquist2016,teixeira2021} (only $\epsilon_{AC} \ne 0$). c) and d):
chains of linkers, branched when bonded to particles. c) competition between branching and chaining (only $\epsilon_{AA}$ and $\epsilon_{AC} \ne 0$) - model A. d): no competition between branching and chaining (only $\epsilon_{AA}$ and $\epsilon_{BC} \ne 0$) - model B.}
\label{fig:structures}
\end{figure}

The interaction potential defined in (\ref{potential}) depends on the number of patches of different types on each species ($f_C$ and $f_B$) and on the energy scales $\epsilon_{\alpha\beta}$ between all pairs of types of patches. The choice of $f_C$, $f_B$ and of the non-zero $\epsilon_{\alpha\beta}$ (i.e. of which types of bonds can be formed) will define a particular model of a binary mixture of patchy particles. The types of bonds allowed will then determine the types of self-assembled structures that emerge on the model. This relation is illustrated in Fig.~\ref{fig:structures}. 
If the only non-zero energy is $\epsilon_{AA}$ then only bonds between linkers can be  formed, and chains of linkers are the only self-assembled structures expected - Fig.~\ref{fig:structures}a). The formation of clusters where  a single linker mediates the bonds between two (and only two) particles \cite{lindquist2016,teixeira2021} is obtained for a model where only  $\epsilon_{AC}\ne 0$ (or only $\epsilon_{AB}\ne 0$ and $f_B=2$) - Fig.~\ref{fig:structures}b). 

The goal of this work is to study the percolation thresholds in systems where the linkers, besides mediating the bonding of particles, can also self-assemble into chains. This will be accomplished by investigating the following models:
\begin{description}
    \item [Model A]{only $AA$ and $AC$ bonds can form, i.e. only $\epsilon_{AC}$ and $\epsilon_{AA}$ are non-zero; in simulations (and numerical calculations) $f_C=4$, and several values of the energy ratio $\epsilon_{AA}/\epsilon_{AC}$ are considered.}
    \item[Model B]
    {only $AA$ and $BC$ bonds can form, i.e. only $\epsilon_{BC}$ and $\epsilon_{AA}$ are non-zero; in simulations (and numerical calculations) $f_C=4$, $f_B=2$,   and several values of the energy ratio $\epsilon_{AA}/\epsilon_{BC}$ are considered.}
\end{description}
In both models,  particles and linkers self-assemble into structures formed by chains of linkers (sequences of $AA$ bonds) that may branch when bonded to particles (through $AC$ bonds in model A or $BC$ bonds in model B - see Figs.~\ref{fig:structures}c) and d)). However, the interplay between chaining and branching is different: in model A, a patch A can bond to patches A  or to patches C, setting a competition between chaining and branching (see Fig.~\ref{fig:structures}c)); in model B, each type of patch is only engaged in one type of bond and this competition is absent (see Fig.~\ref{fig:structures}d)). 

It is worthwhile noticing that model A and model B (with $f_B=2$) become the thoroughly studied linker particle model of \cite{lindquist2016,teixeira2021} when $\epsilon_{AA}=0$. Therefore, the present work can be seen as an extension of the study of linker-particle  models to a case where  the self-assembly of linkers into chains is present. 



\subsection{\label{sec:simulations}Simulations}

Structural properties of the models were obtained as a function of  temperature, density, and composition with a classic Metropolis Monte Carlo (MC) simulation in the canonical ensemble, using \cite{sciortino2007} as reference.
A mixture of 1200 particles and linkers in a cubic box with periodic boundary conditions was set up. Particles and linkers were randomly placed in the box with random orientations, and then moved around until equilibrium was reached. A move was defined as a simultaneous displacement between $\pm0.05\sigma$ in each direction and a rotation between $\pm0.1$ rad around a random axis, with all quantities being drawn from uniform distributions.  Simulations ran for a minimum of $10^5$ MC steps (each step is defined as 50 000 attempts to move a particle or a linker). Steadiness of the bonding probabilities (i.e. of the number of bonds formed in the system) and the fraction of particles belonging to the largest cluster were used to assess equilibrium. Cluster-size distributions were obtained using the Hoshen-Kopelman algorithm \cite{Hoshen1976}. 

The particles were decorated with $f_C=4$ patches of type C, placed on their surface as vertices of a  tetrahedron. The linkers were decorated with 2 patches of type A 
as (opposite) poles, and 
with $f_B=2$ patches of type B, placed at the equator and as opposite poles. 
The radius of the patches (or the range of the patch-patch potential) 
was set to $\delta=(\sqrt{5-2\sqrt{3}}-1)\sigma/2\approx0.119\sigma$, the maximum value that guarantees that each patch is engaged in a single bond.

\subsection{\label{sec:theory}Theory}

The percolation thresholds are calculated using a generalized Flory-Stockmayer random-bond percolation theory \cite{flory1941,stockmayer1943,tavares2010} for mixtures with several types of bonds \cite{delasHeras2011b,teixeira2021}. In this theory, closed loops are neglected and the clusters assume a tree-like bonding structure. As a consequence, the particles of a cluster can be separated by levels: a random particle is chosen as the level 0; the particles bonded to this are at level 1, and so-forth. This approach is briefly reviewed here for the case in which patches of a given type are only present in one of the species of the mixture, as happens in the models under study. The number of patches of type $\gamma$ that belong to particles of level $i$ and are  bonded to particles of the previous level, $b_{i,\gamma}$, 
follows a recursive relation that can be expressed in a matrix form,
\begin{equation}
    \label{matrix}
    \tilde b_i={\tilde T}^i \tilde b_1,
\end{equation}
where $\tilde b_i$ is a vector with components $b_{i,\gamma}$. The matrix $\tilde T$ encodes the structure of the clusters and is a function of $f_ \gamma$, the number of patches of type $\gamma$ in a particle, and of the probabilities $p_{\alpha \to \beta}$ that are the fraction of patches of type $\alpha$ that are bonded to patches of type $\beta$.
Percolation will be absent if the absolute values of all eigenvalues $\lambda$ of $\tilde T$ are lower than one. The percolation threshold is obtained, as a  function of $f_\gamma$ and of $p_{\alpha \to \beta}$, by finding the conditions 
for which 1 is the largest absolute value of all eigenvalues of matrix $\tilde T$.

The probabilities $p_{\alpha \to \beta}$ are obtained as a function of density, composition and temperature, 
  through the laws of mass action of Wertheim's theory  \cite{delasHeras2011b,tavares2010} that provide a connection between  percolation and thermodynamics.  
Formation of bonds between patches of type $\alpha$ and patches of type $\beta$ can be seen as a chemical reaction whose equilibrium is obtained when,
\begin{equation}
\label{genlma}
p_{\alpha \to \beta}=x_{\beta}\rho (1-p_{\alpha}) (1-p_{\beta})
f_{\beta} \Delta_{\alpha\beta},
\end{equation}
where $x_\beta$ is the fraction of particles that contain patches of type $\beta$ and $p_\alpha\equiv \sum_{\gamma} p_{\alpha \to \gamma}$ is the fraction of patches of type $\alpha$ that are bonded. $\Delta_{\alpha\beta}$ is interpreted as being the reaction constant for the formation of bonds $(\alpha\beta)$ \cite{tavares2010} or the partition function of these bonds \cite{bianchi2008}. For the interaction potential  (\ref{potential}), $\Delta_{\alpha\beta}$   is obtained using
 Wertheim's first order perturbation theory  and a linear approximation for the HS fluid pair correlation function \cite{bianchi2008}:
 \begin{equation}
 \label{Deltaalphabeta}
     \Delta_{\alpha\beta}=\frac{v_{\alpha\beta}}{\sigma^3}\left[\exp\left(\frac{\epsilon_{\alpha\beta}}{k_BT}\right)-1\right] G(\frac{\pi}{6}\rho,\frac{\delta_{\alpha\beta}}{\sigma}),
 \end{equation}
where $k_B$ is Boltzmann's constant, $T$ is the temperature, 
\begin{equation}
    \label{Gxy}
    G(z,y)=\frac{1}{(1-z)^3}\left(1-\frac{5}{2}\frac{3+8y+3y^2}{15+4y}z-
  \frac{3y}{2}\frac{1+5y}{15+4y}z^2\right),  
\end{equation}
and,
\begin{equation}
\label{vb}
    v_{\alpha\beta}=\frac{\pi}{30}\frac{\delta}{\sigma}\left(15+4\frac{\delta}{\sigma}\right)\delta^3.
\end{equation}
$v_{\alpha\beta}$ is the bonding volume, i.e. the volume that can be explored by one particle when keeping the other particle fixed, without breaking the bond $\alpha\beta$ between the two. In most calculations $\delta_{\alpha\beta}$ will be equal to the value used in simulations (i.e. $\delta_{\alpha\beta}=\delta=0.119\sigma$).
The calculation of the probabilities $p_{\alpha \to \beta}$ as a function of thermodynamic quantities is completed by using the normalization $p_\alpha=
\sum_\beta p_{\alpha \to \beta}$ in (\ref{genlma}):
\begin{equation}
\label{lma}
    p_\alpha=\rho(1-p_\alpha)\sum_\beta f_\beta x_\beta (1-p_\beta) \Delta_{\alpha\beta}.
\end{equation}
These equations (whose number equals the total number of different types of patches) are the laws of mass action. By solving them the bonding probabilities $p_\alpha$ are obtained from the thermodynamic quantities, and the probabilities $p_{\alpha \to \beta}$ can then be computed from (\ref{genlma}) and introduced in matrix $\tilde T$ of (\ref{matrix}). Finally, the eigenvalues $\lambda$ are obtained, and  the percolation threshold is determined as a function of $\rho$, $T$ and $x$ for a given model.

 In what follows we present the specific expressions of  matrix $\tilde T$ and of equations (\ref{genlma}) and (\ref{lma}) for  models A and B.

 \subsubsection{Model A}
In model A, linkers have 2 patches of type $A$ and particles have $f_C$ patches of type $C$.  Only  bonds $AA$ and bonds $AC$ can form; as a consequence, matrix $\tilde T$ is (see appendix \ref{appMatrix}),  
\begin{equation}
\label{matrixA}
\tilde T =\begin{bmatrix}
      p_{A\to A} & (f_C-1) p_{C\to A}  \\
       p_{A\to C} & 0
    \end{bmatrix}.
\end{equation}
This matrix has two eigenvalues, one positive, $\lambda_+$, and one negative $\lambda_-$, with $\lambda_+ >|\lambda_-|$. The equation for the percolation threshold is obtained by imposing  $\lambda=1$ as a root of the characteristic polynomial of (\ref{matrixA}),
\begin{equation}
\label{percmodA}
(1-p_{A\to A})=(f_C-1)p_{C\to A} p_{A\to C}.
\end{equation}
The probabilities $p_{\alpha \to \beta}$ are related to the bonding probabilities $p_\alpha$ using (\ref{genlma}) (recall that composition $x$ is the fraction of particles to which patches of type $C$ belong):
\begin{equation}
\label{pAAmodA}
p_{A\to A}=2(1-x)\rho (1-p_A)^2 \Delta_{AA},
\end{equation}
\begin{equation}
\label{pACmodA}
p_{A\to C}=f_Cx\rho (1-p_A) (1-p_C) \Delta_{AC},
\end{equation}
and
\begin{equation}
\label{pCAmodA}
p_{C\to A}= \frac{2 (1-x)}{f_C x}p_{A\to C}.
\end{equation}
The laws of mass action are,
\begin{equation}
\label{lma1}
p_A= p_{A\to A}+p_{A\to C},
\end{equation}
and,
\begin{equation}
\label{lma2}
p_C = p_{C\to A}.
\end{equation}
The results of previous works where no $AA$ bonds were allowed \cite{lindquist2016,teixeira2021} are recovered when $p_{A\to A}\equiv 0$. 

\subsubsection{Model B}
In model B, linkers have 2 patches of type $A$ and $f_B$ patches of type B, while  particles have $f_C$ patches of type $C$. Only  bonds $AA$ and  bonds $BC$ can form; as a consequence, matrix $\tilde T$ is (see appendix \ref{appMatrix}),  
\begin{equation}
\label{matrixB}
\tilde T= \begin{bmatrix}
      p_{A\to A} & 2 p_{A\to A} & 0 \\
      0 & 0 & (f_C-1)p_{C\to B} \\
      f_B p_{B\to C} & (f_B-1) p_{B\to C} & 0
    \end{bmatrix}.
\end{equation}
The equation for the percolation threshold is obtained by imposing  $\lambda=1$ as a root of the characteristic polynomial of (\ref{matrixB}) (see appendix \ref{apppercB}),
\begin{equation}
\label{percmodB}
1-p_{A\to A}=(f_C-1)p_{C\to B} p_{B\to C}\left[(f_B+1) p_{A\to A}+(f_B-1)\right].
\end{equation}
The probabilities $p_{\alpha \to \beta}$ are related to the bonding probabilities $p_\alpha$ using (\ref{genlma}):
\begin{equation}
p_{A\to A}=2(1-x)\rho (1-p_A)^2 \Delta_{AA},
\end{equation}
\begin{equation}
p_{B\to C}=f_C x\rho (1-p_B) (1-p_C) \Delta_{BC},
\end{equation}
and
\begin{equation}
p_{C\to B}= \frac{f_B (1-x)}{f_C x}p_{B\to C}.
\end{equation}
The laws of mass action are,
\begin{equation}
p_A = p_{A\to A},
\end{equation}
and,
\begin{equation}
p_C = p_{C\to B}.
\end{equation}

\section{\label{sec:results}Results}

\begin{figure}[htb]
\advance\leftskip-0.1cm
\begin{overpic}[width=1.02\columnwidth]{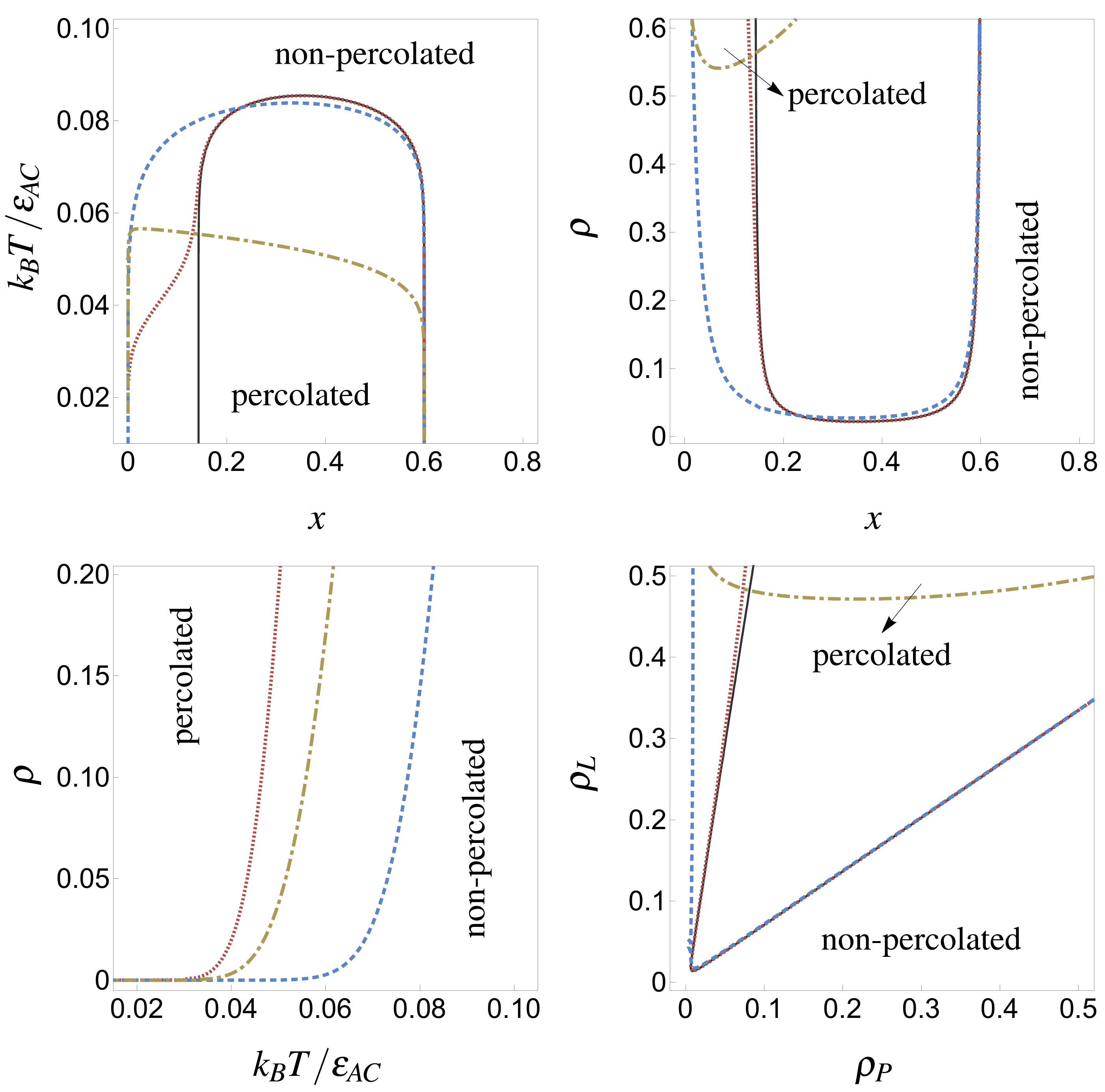}
 \put (0.5,95) {a)}
 \put (9.5,50) {\protect\includegraphics[width=0.02\textwidth]{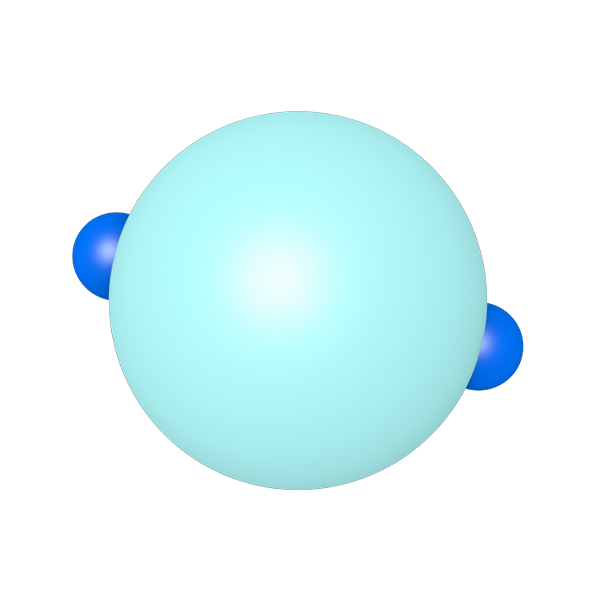}}
 \put (45.5,50) {\protect\includegraphics[width=0.02\textwidth]{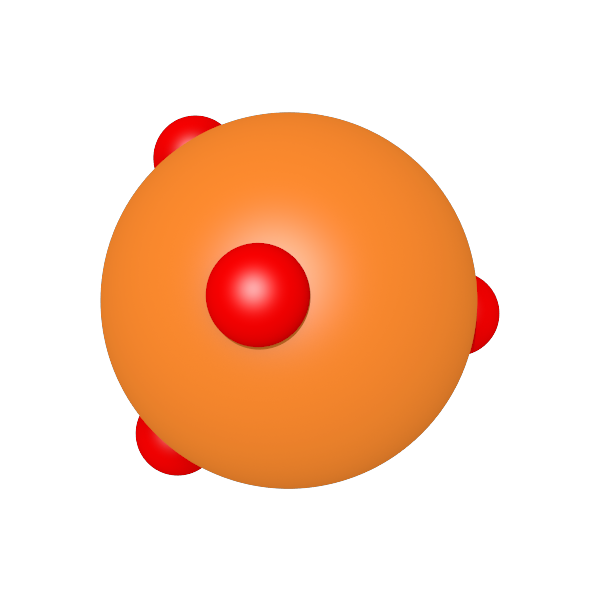}}
 \put (51.2,94.8) {b)}
 \put (60,50) {\protect\includegraphics[width=0.02\textwidth]{images/particles/linkerAC_plots.png}}
  \put (95.8,50) {\protect\includegraphics[width=0.02\textwidth]{images/particles/coll_plots.png}}
 \put (0.5,45.5) {c)}
 \put (51.5,45.2) {d)}
\end{overpic}
\caption{Percolation thresholds for model A, with $f_C=4$ and the following values of  $\epsilon_{AA}/\epsilon_{AC}$: 0 (black full line), \, 0.5 (red dotted line), \, 1.0 (blue dashed line), \, 1.5 (green dot-dashed line).  Percolation occurs in the indicated regions of the diagrams.
(a) Temperature-composition percolation diagram at density $\rho=0.1$; percolation occurs below the lines. 
(b) Density-composition percolation diagram at temperature  $k_BT/\epsilon_{AC}=0.075$; percolation occurs above the lines; (c) Density-temperature percolation diagram at composition $x=0.1$; percolation occurs above the lines. (d) density of linkers, $\rho_L\equiv (1-x)\rho$, vs density of particles, $\rho_P\equiv x\rho$, percolation diagram at fixed temperature $k_BT/\epsilon_{AC}=0.075$; percolation occurs inside the region limited by the lines.}  
\label{fig:percmodA1}
\end{figure}

\subsection{Model A with $\epsilon_{AA}/\epsilon_{AC}<2$}

The percolation diagram of model A when linkers do not form chains (i.e when $\epsilon_{AA}=0$) has already been obtained \cite{lindquist2016,teixeira2021} and is represented in Fig.~\ref{fig:percmodA1} (full line in all panels). 
Percolation is present only for a limited range of compositions, $ 1/7 \le x \le 0.6$, at low temperatures
 - Fig.~\ref{fig:percmodA1}a). These two limiting compositions can be determined from (\ref{percmodA},\ref{pACmodA}) with $p_{A\to A}=0$ (and $f_C=4$): when $p_{C\to A} \to 1$ (i.e. all patches C are bonded), $x\to 1/7$; when $p_{A\to C} \to 1$ (i.e. all patches A are bonded), $x\to 0.6$. This has a transparent physical meaning: when chains of linkers are absent, the system needs both a minimum amount of particles ($1/7$) and a minimum amount of linkers ($0.4$) to form large clusters. Below $x=1/7$, there are too many linkers: part of them bond to patches $C$, almost fully covering the particles and preventing significant  bonding between two particles (patches $C$ are "blocked"); the other linkers are free and do not contribute to clustering. Above $x=0.6$ the linkers can promote the bonding of particles, but their number is not enough to form  large clusters. 
\begin{figure}[htb]
\advance\leftskip-0.1cm
\begin{overpic}[width=1.02\columnwidth]{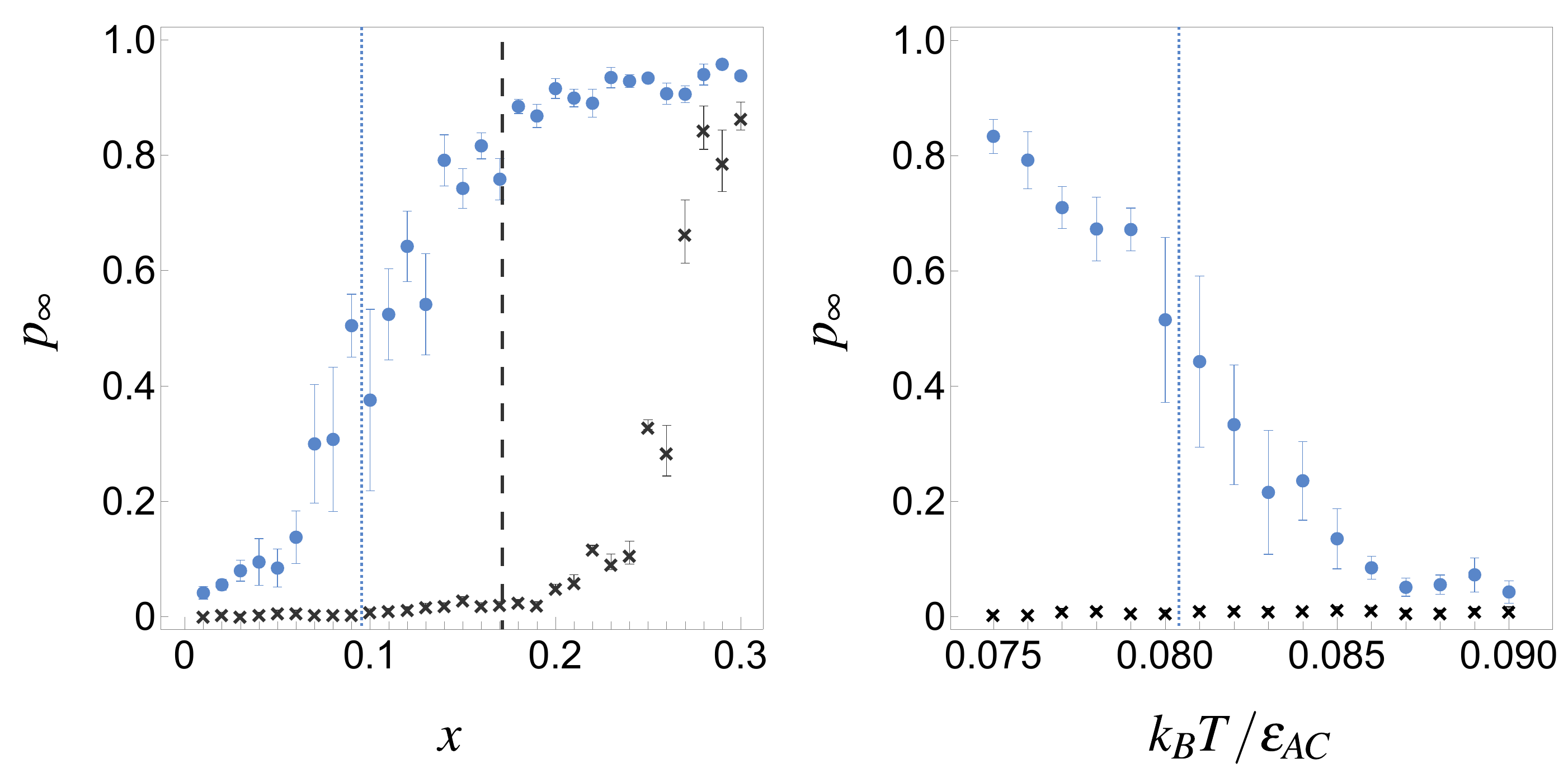}
 \put (1.5,45.5) {a)}
 \put (10,0.5) {\protect\includegraphics[width=0.02\textwidth]{images/particles/linkerAC_plots.png}}
  \put (45.5,0.5) {\protect\includegraphics[width=0.02\textwidth]{images/particles/coll_plots.png}}
 \put (51,45) {b)}
\end{overpic}
\caption{ 
Fraction of particles and linkers that belong to the largest cluster, $p_\infty$, obtained from simulations for model A. Blue circles and black crosses correspond to $\epsilon_{AA}/\epsilon_{AC}=1$ and
$\epsilon_{AA}/\epsilon_{AC}=0$, respectively: (a) 
$\rho=0.15$ and $k_BT/\epsilon_{AC}=0.08$; (b) $\rho=0.15$ and $x=0.1$.  The vertical lines signal the composition (in (a)) and the temperature (in (b)) at which the theory predicts percolation to occur for $\epsilon_{AA}/\epsilon_{AC}=1$ (blue dotted line) and for $\epsilon_{AA}/\epsilon_{AC}=0$ (black dashed line).    
}  
\label{fig:simsmodA1}
\end{figure}
The introduction of chains of linkers changes qualitatively the temperature-composition percolation diagram for $x\le 0.6$ - see Fig.~\ref{fig:percmodA1}a). In particular, the limit $x=1/7$  disappears and percolation can be obtained for all compositions up to $x=0.6$ at low temperatures. The cause of this change is energy minimization. 
A bond $AC$ decreases the energy by $\epsilon_{AC}$; the formation of 
a bond $AA$ from bonds $AC$ requires the breaking of two bonds $AC$ (to free two patches $A$), and the energy variation resulting from this process is $-\epsilon_{AA}+2\epsilon_{AC}$. As a consequence, at low temperatures and when $\epsilon_{AA}<2\epsilon_{AC}$, patches $A$ will bond preferentially to patches $C$. At low composition $x$, bonds $AC$ will saturate patches $C$, but all the remaining patches $A$ can now bond to form $AA$ bonds. Therefore, the energy is minimized by  forming chains of linkers ($AA$ bonds) that branch when  connected to particles ($AC$ bonds). This structure is percolated and will always form for $0<\epsilon_{AA}/\epsilon_{AC}<2$ and $x<0.6$ at sufficiently low temperatures.
Therefore, the self-assembly of linkers into chains opens the possibility of percolation with fewer particles (for the same amount of linkers).

These theoretical predictions are confirmed with simulations, as shown in Fig.~\ref{fig:simsmodA1}. 
For a series of equilibrated simulations at fixed $(x,\rho,T)$, the cluster size distribution was recorded at each $10^3$ MC steps, the fraction of particles and linkers that belong to the largest cluster was determined, and its mean value (the points in Figs.~\ref{fig:simsmodA1} and \ref{fig:simsmodB1}) and standard deviation (the error bars in Figs.~\ref{fig:simsmodA1} and \ref{fig:simsmodB1}) were calculated.  
The fraction of particles and linkers that belong to the largest cluster, $p_\infty$, is displayed, for $\epsilon_{AA}/\epsilon_{AC}=0$ and $\epsilon_{AA}/\epsilon_{AC}=1$ at a fixed low density, as a function of $x$ for a given low temperature in Fig.~\ref{fig:simsmodA1} a), and as a function of temperature for a given low  $x$ in Fig.~\ref{fig:simsmodA1}b). $p_\infty$ is the order parameter for the percolation transitions, going from 0 (in the thermodynamic limit) below the percolation threshold to a non-zero value  above the percolation threshold.  For the finite systems employed, the sudden increase of $p_\infty$ signals the onset of percolation. 
The simulation results of Fig.~\ref{fig:simsmodA1}a) confirm that the formation of chains of linkers decreases the fraction of particles needed to obtain percolation: the strong variation of $p_\infty$ is observed at lower $x$  for $\epsilon_{AA}/\epsilon_{AC}=1$. On the other hand, the results depicted in Fig.~\ref{fig:simsmodA1}b), show that for a fixed low $x$, percolation is only obtained in the case $\epsilon_{AA}/\epsilon_{AC}=1$.  


\begin{figure}[thb]
\includegraphics[width=0.85\columnwidth]{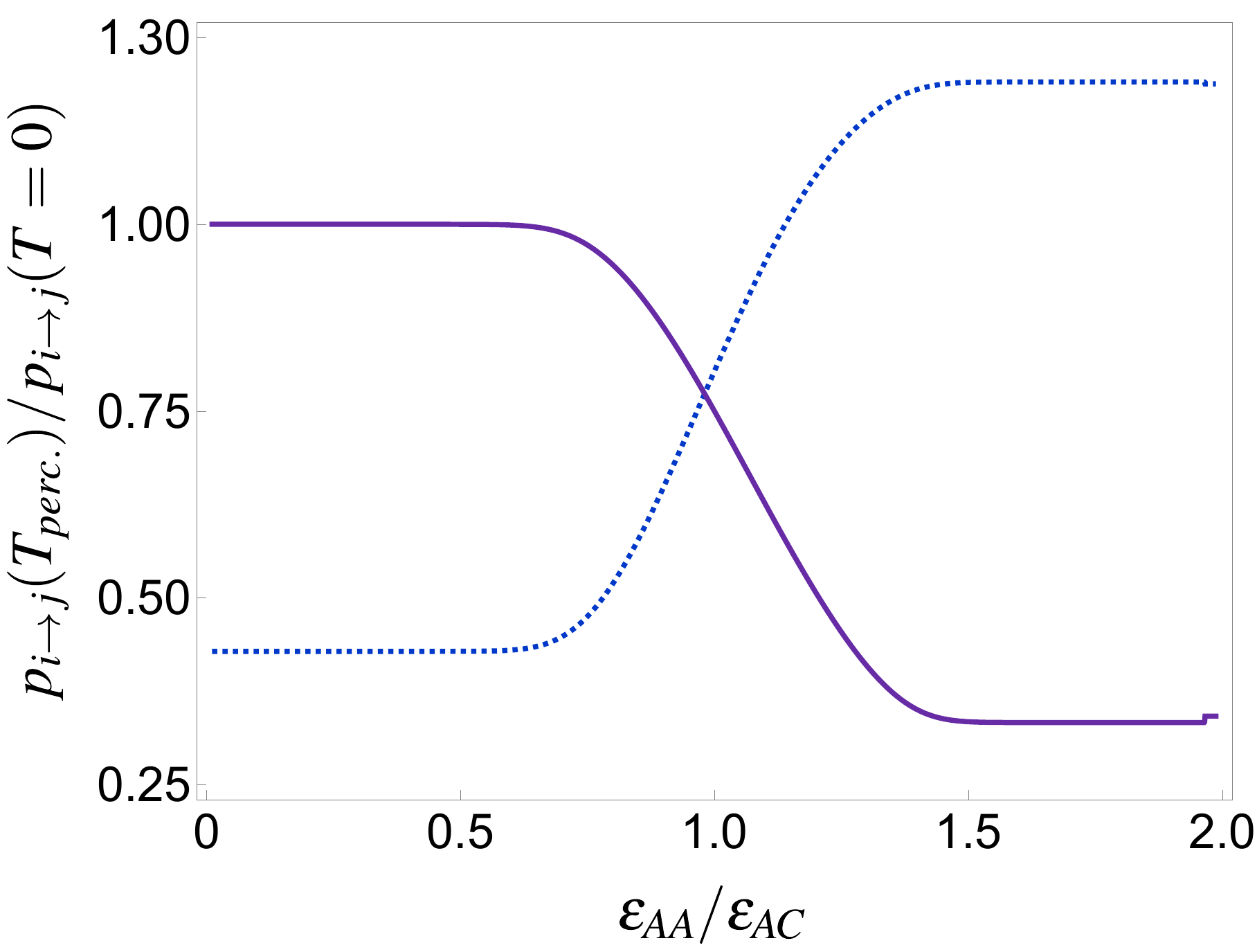}
\caption{ Ratio between  $p_{A\to A}$ calculated at the percolation threshold and $p_{A\to A}$ calculated at $T=0$ (blue dotted line) 
and ratio between  $p_{C\rightarrow A}$ calculated at the percolation threshold and $p_{C\to A}$ calculated at $T=0$ (purple full line), as a function of $\epsilon_{AA}/\epsilon_{AC}$ for fixed $(\rho,x)=(0.1,0.1)$.   These quantities represent the fraction of $AA$ bonds and of $AC$ bonds, respectively, that do not break when a system is heated from $T=0$ to the temperature of the percolation threshold, $T_{perc}$. The cases in which $p_{A\rightarrow A}(T_{perc})/p_{A\rightarrow A}(T=0)>1$ mean that $AA$ bonds were formed during this heating process.}
\label{fig:pijeAAeAC}
\end{figure}

The temperature at which, for low $x$, percolation disappears, has a non-monotonic dependence on $\epsilon_{AA}/\epsilon_{AC}$, as can be seen if Figs.~\ref{fig:percmodA1}a) and c).  This can be understood by recognizing that there are three ways  of disrupting the fully bonded percolated network that is formed at low temperatures: (a) breaking the chains (or $AA$ bonds) that connect the particles; (b) breaking the bonds between the chains and the particles (or $AC$ bonds); (c) a combination of (a) and (b). 
These three regimes can be  identified by determining the fraction of bonds $AA$ and $AC$ that do not break when the system  is heated from $T=0$ to the percolation temperature. These quantities (calculated at $(\rho,x)=(0.1,0.1)$) are represented, as a function of $\epsilon_{AA}/\epsilon_{AC}$, in Fig.~\ref{fig:pijeAAeAC}:
\begin{itemize}
\item[(a)]{For low values of $\epsilon_{AA}/\epsilon_{AC}$  (up to $0.5$) percolation disappears due to the breaking of chains: around $60\%$ of bonds $AA$ are broken, while $AC$ bonds remain unchanged. In this case, the energy cost of breaking $AA$ bonds is low, and so the breaking of chains in an extent that leads to the disappearance of percolation may happen at a low temperature.}
\item[(b)]{For the larger values of $\epsilon_{AA}/\epsilon_{AC}$ (in the range $\approx 1.5$ to 2), the network is disrupted due to the breaking of the bonds between chains and particles: $2/3$ of the bonds $AC$ break, while new bonds $AA$ are formed. In this regime, the energy cost of replacing  2 $AC$ bonds by one $AA$ bond and 2 free patches $C$ is low, and therefore the percolation threshold happens at low temperatures.}
\item[(c)] {For intermediate values of $\epsilon_{AA}/\epsilon_{AC}$, percolation disappears through both mechanisms: when $\epsilon_{AA}=\epsilon_{AC}$, around $25\%$ of bonds $AA$ and bonds $AC$ are broken at the percolation threshold. In this regime, the energy cost of breaking chains is larger than in (a) and the energy cost of replacing bonds $AC$ by bonds $AA$ and 2 free $C$ patches is larger than in (c). As a consequence the temperature at which percolation disappears is larger than in those regimes.}
\end{itemize}

The dependence of the percolation threshold on density is illustrated in Figs.~\ref{fig:percmodA1}b), c) and d). For a fixed intermediate temperature ($k_BT/\epsilon_{AC}=0.075$ in Fig.~\ref{fig:percmodA1}b))  there is percolation above a minimum density, for a range of compositions $x$.  For values of $\epsilon_{AA}/\epsilon_{AC}$ up to 1 this minimum density is low and doesn't change;  the lower limit of the range of compositions, on the other hand, decreases significantly when $\epsilon_{AA}/\epsilon_{AC} \approx 1$.  This same effect is evident in Fig.~\ref{fig:percmodA1}d): the minimum number of particles needed to obtain percolation for a fixed number of linkers decreases significantly for those values of $\epsilon_{AA}/\epsilon_{AC}$. On the contrary, the maximum number of particles for which, for a given number of linkers, percolation still exists, is unaffected by $\epsilon_{AA}/\epsilon_{AC}$. This behavior is totally different for   $\epsilon_{AA}/\epsilon_{AC}=1.5$:  at $k_BT=0.075$ percolation is obtained  only at high densities (Figs.~\ref{fig:percmodA1}b) and d));   percolation takes place at low densities only if the temperature is significantly lowered - see Fig.~\ref{fig:percmodA1}c).

\subsection{Model A with $\epsilon_{AA}/\epsilon_{AC}>2$.} 

The structure of the ground state of model A changes  when $\epsilon_{AA}> 2\epsilon_{AC}$. It becomes energetically more favourable to form bonds $AA$  than bonds $AC$, and all linkers tend to assemble into long chains without connecting to the particles. Only chains with no branching are formed at low temperatures and  so percolation is not energetically favourable. We found no numerical solutions to the equations (\ref{percmodA},\ref{pAAmodA},\ref{pACmodA},\ref{pCAmodA},\ref{lma1},\ref{lma2}) using $\delta_{AC}=\delta_{AA}=\delta=0.119\sigma$  and $\epsilon_{AA} >2\epsilon_{AC}$.

\begin{figure}[htb]
\advance\leftskip-0.1cm
\begin{overpic}[width=1.02\columnwidth]{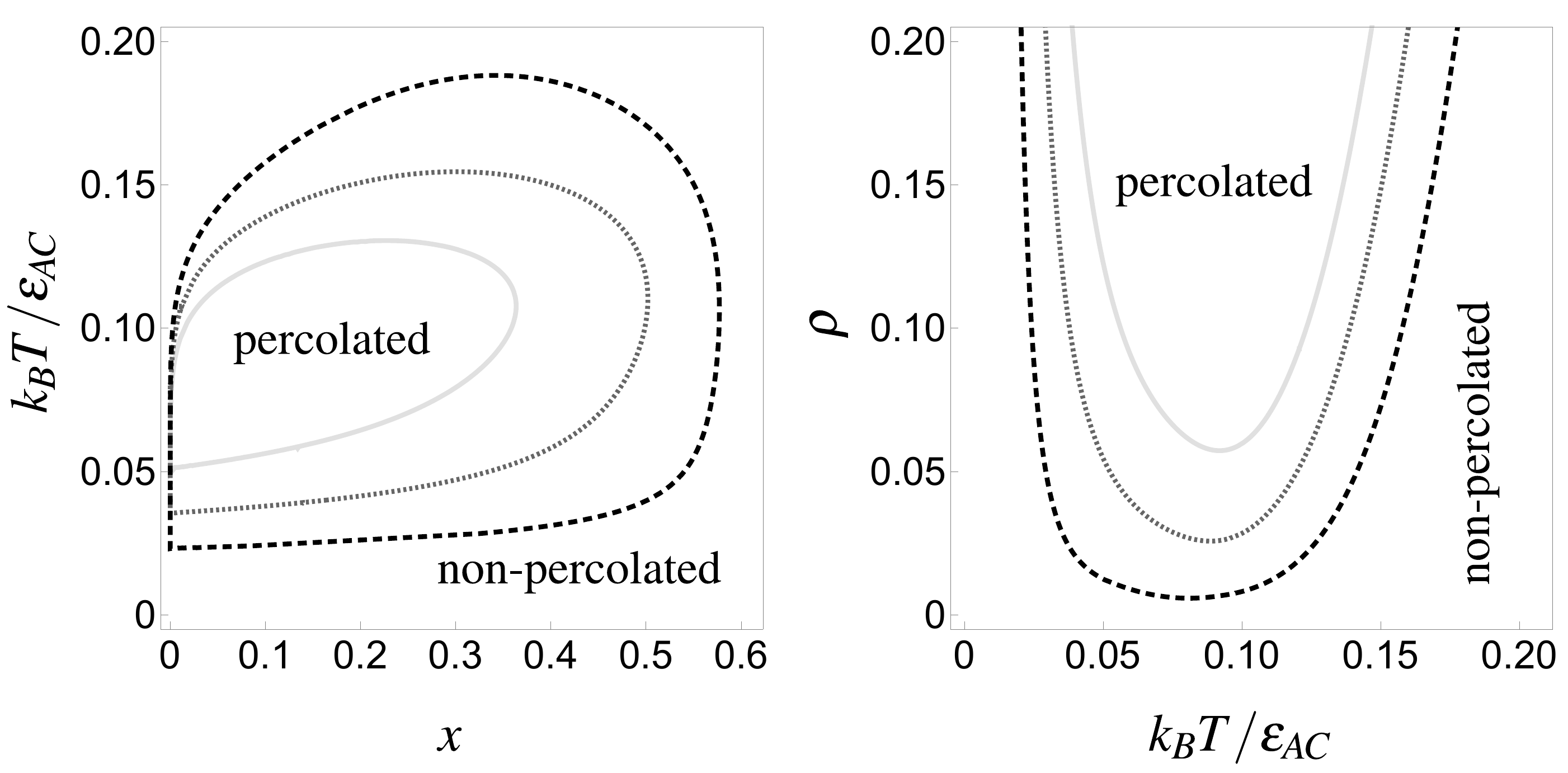}
 \put (0.5,45.5) {a)}
 \put (10,0.2) {\protect\includegraphics[width=0.02\textwidth]{images/particles/linkerAC_plots.png}}
  \put (45.5,0.2) {\protect\includegraphics[width=0.02\textwidth]{images/particles/coll_plots.png}}
 \put (51,45) {b)}
\end{overpic}
\caption{Percolation thresholds for $\delta_{AC}=$0.45 (full line), 0.5 (dotted line), 0.6 (dashed line), when $\epsilon_{AA}/\epsilon_{AC}=2.1$. (a) temperature- composition diagram at $\rho=0.1$; (b) density-temperature diagram at $x=0.1$.}  
\label{fig:eAA21}
\end{figure}

However, it is possible to envisage a mechanism that promotes percolation when it is not energetically favourable to form branched structures. As temperature is raised, bonds $AA$ start to break and to free $A$ patches that can bond to $C$ patches.
The resulting structure will be determined by the competition between these two types of entropic defects \cite{tavares2009,russo2011}:
free patches $A$ and bonds $AC$. Favouring bonds $AC$ promotes branching and the possibility of percolation. 
These bonds can be made entropically more favourable  by increasing  their volume $v_{AC}$.  Figure \ref{fig:eAA21} shows the percolation thresholds for $\epsilon_{AA}/\epsilon_{AC}=2.1$ and values of $\delta_{AC}/\sigma=0.45,\,0.5,\, 0.6$ (i.e. larger values of the bonding volume $v_{AC}$). $\delta_{AA}$ is kept equal to $0.119\sigma$. The temperature-composition diagram of Fig.~\ref{fig:eAA21}a) exhibits  a percolation loop that appears only when $\delta_{AC} \gtrsim 0.4$ and that increases in size when $\delta_{AC}$ is increased. This result shows that, at fixed low $\rho$ and intermediate $x$, the temperature increase favours, at first, "entropic" branching in an extent that makes percolation possible. Further increase of the temperature promotes breaking of both types of bonds and, at some point, the vanishing of percolation. Therefore,  the non-intuitive phenomena of percolation (or gellation) by heating \cite{roldanvargas2013} is obtained  as a result of an entropic competition in bond formation.
The density of the percolation threshold is represented in Fig.~\ref{fig:eAA21}b) as a function of temperature, at $x=0.1$: there is a density below which percolation is not possible; for larger densities, percolation occurs within a range of temperatures.

The use of $\delta_{AC}>0.119\sigma$ in the theoretical framework developed in section \ref{sec:theory} is, strictly speaking, inconsistent. The approximation for $\Delta_{AC}$ in (\ref{Deltaalphabeta}) and the tree like cluster description employed for the structure both  assume that each patch can only be engaged in  one bond (single bond per patch condition). This hypothesis is violated, for the  interacting potential (\ref{potential}), when $\delta_{\alpha\beta}>0.119\sigma$. However, the results represented in Fig.~\ref{fig:eAA21} are still physically meaningful. In fact, flexible linkers (or particles) and mobile patches on particles, could give origin to larger bonding volumes without compromising the single bond per patch condition.  Therefore, the results of Fig.~\ref{fig:eAA21} suggest that in systems where linear linker-linker and low valence linker-particle self-assembly occur, entropic effects related to the flexibility and mobility of linkers can originate unexpected features like percolation (or gellation) by heating and  closed loops in percolation diagrams.  

\subsection{Model B}

Model B is characterized by the self-assembly of linkers into chains through $AA$ bonds and by the branching of these chains when they connect to the particles through $BC$ bonds. Since each type of patch is only involved in one type of bond, the formation of bonds $AA$ does not constrain the formation of $BC$ bonds (and vice-versa). As a consequence, the structure of the ground state is independent of $\epsilon_{AA}/\epsilon_{BC}$ and corresponds to the maximization of the number of both types of bonds. 

\begin{figure}[htb]
\advance\leftskip-0.1cm
\begin{overpic}[width=1.02\columnwidth]{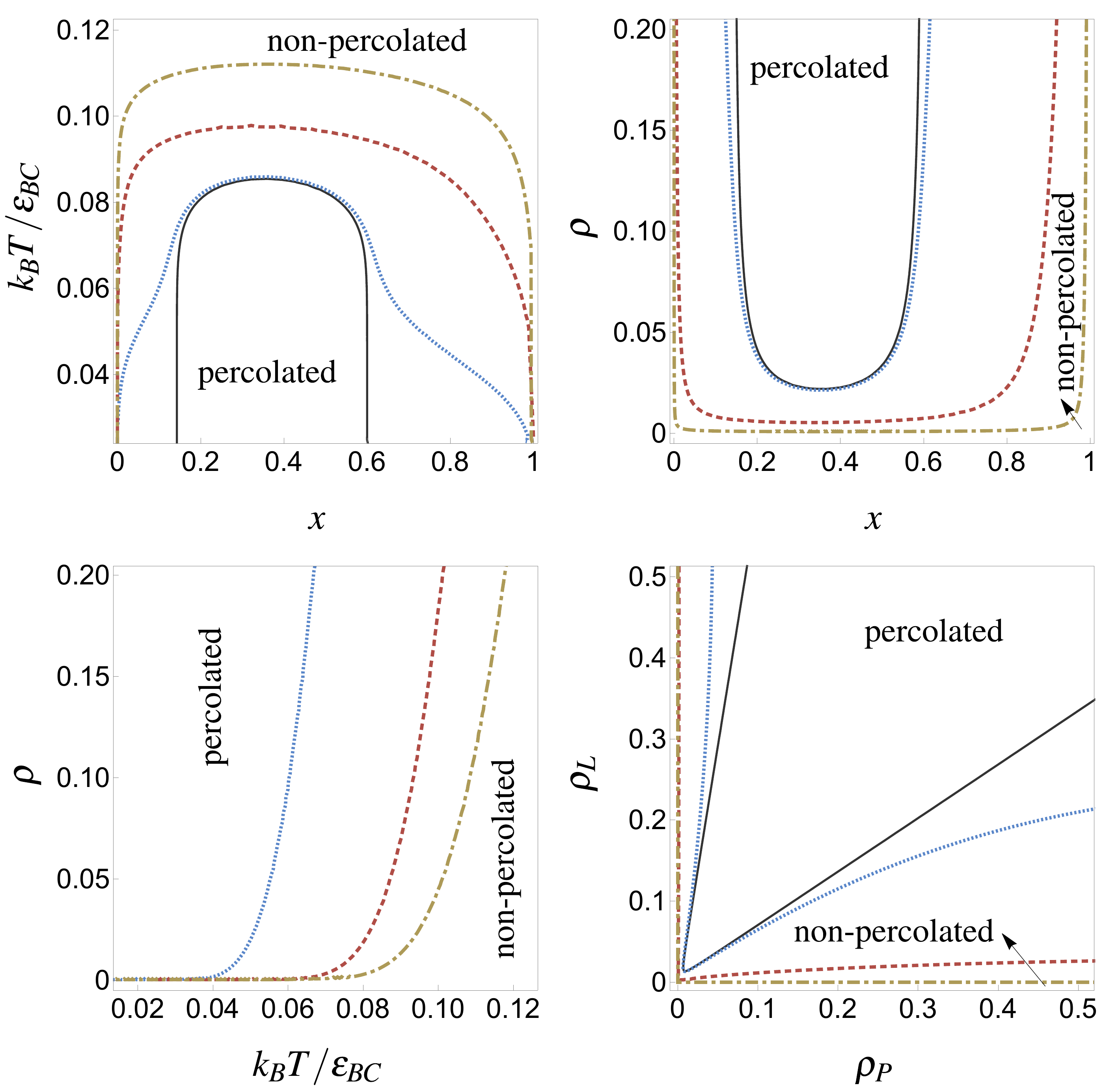}
 \put (0.5,95.1) {a)}
 \put (9.5,50) {\protect\includegraphics[width=0.02\textwidth]{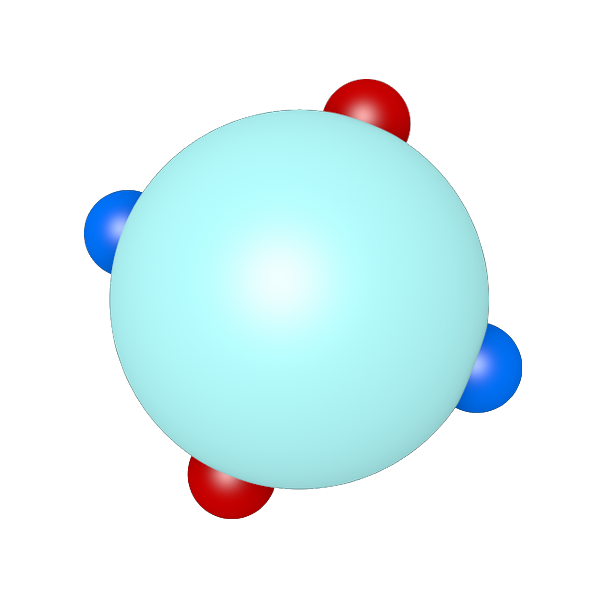}}
 \put (45.5,50) {\protect\includegraphics[width=0.02\textwidth]{images/particles/coll_plots.png}}
 \put (51.2,94.8) {b)}
 \put (60,50) {\protect\includegraphics[width=0.02\textwidth]{images/particles/linkerAABB_plots.png}}
  \put (95.8,50) {\protect\includegraphics[width=0.02\textwidth]{images/particles/coll_plots.png}}
 \put (0.5,45.4) {c)}
 \put (51.5,45.2) {d)}
\end{overpic}
\caption{Percolation thresholds for model B, with $f_C=4$, $f_B=2$ and the following values of  $\epsilon_{AA}/\epsilon_{BC}$: 0 (black full line), \, 0.5 (blue dotted line), \, 1.0 (red dashed line), \, 1.5 (green dot-dashed line).  Percolation occurs in the indicated regions of the diagrams.
(a) Temperature-composition percolation diagram at density $\rho=0.1$; percolation occurs below the lines. 
(b) Density-composition percolation diagram at temperature  $k_BT/\epsilon_{BC}=0.075$; percolation occurs above the lines; (c) Density-temperature percolation diagram at composition $x=0.1$; percolation occurs above the lines. (d) density of linkers, $\rho_L\equiv (1-x)\rho$, vs density of particles, $\rho_P\equiv x\rho$, percolation diagram at fixed temperature $k_BT/\epsilon_{BC}=0.075$; percolation occurs inside the region limited by the lines.
}  
\label{fig:percmodB1}
\end{figure}

The percolation diagrams of Fig.~\ref{fig:percmodB1} (for $f_B=2$) demonstrate that the introduction of chain assembly in  model B leads to substantial changes. The temperature-composition diagram of Fig.~\ref{fig:percmodB1}a) shows that percolation is obtained for all compositions $x$ at sufficiently low temperatures. This is a drastic change from the case $\epsilon_{AA}=0$ and $f_C=4$, where percolation could only be found for $1/7<x<0.6$, and from model $A$, where percolation was never found for $x>0.6$. The presence of percolation at extremely low $x$ and at low temperatures means that a vanishing small amount of particles is enough to connect, through $BC$ bonds, the abundant and long chains of linkers formed by $AA$  bonds, in an extent that leads to the formation of percolating clusters. On the other hand, the emergence of percolation at $x$ close to 1 and low temperatures, means that a vanishing small amount of linkers is enough to connect the particles in a percolated cluster, through the combination of a few bonds $BC$ that connect linkers to particles, and of bonds $AA$ that connect linkers to linkers and, indirectly, particles to particles.

The density dependence of percolation is shown in Figs.~
 \ref{fig:percmodB1}b), c) and d). At a given temperature ($k_BT/\epsilon_{BC}=0.075$ in Fig.~\ref{fig:percmodB1}b)), percolation only exists above a certain density, and this minimum density decreases with $\epsilon_{AA}/\epsilon_{BC}$. Above this density, percolation occurs for a range of compositions that increases   with increasing $\epsilon_{AA}/\epsilon_{BC}$. In practice, when $\epsilon_{AA}/\epsilon_{BC}$ is high, percolation occurs for every composition. At fixed composition ($x=0.1$ in Fig.~\ref{fig:percmodB1}c)), the temperature at which percolation occurs for a given density, increases with $\epsilon_{AA}/\epsilon_{BC}$. 
 At a fixed temperature, the density of particles (linkers)  needed to obtain percolation for a given density of linkers (particles), decreases significantly with increasing $\epsilon_{AA}/\epsilon_{BC}$ - see Fig.~\ref{fig:percmodB1}d). 
The results of Fig.~\ref{fig:percmodB1} show that, in model B, when chaining of linkers is favoured, percolation is controlled mainly by temperature (see the results for $\epsilon_{AA}/\epsilon_{BC}=1.5$): below a threshold  temperature (almost constant) percolation occurs for all compositions and all densities above a very small minimum density.

\begin{figure}[htb]
\advance\leftskip-0.1cm
\begin{overpic}[width=1.02\columnwidth]{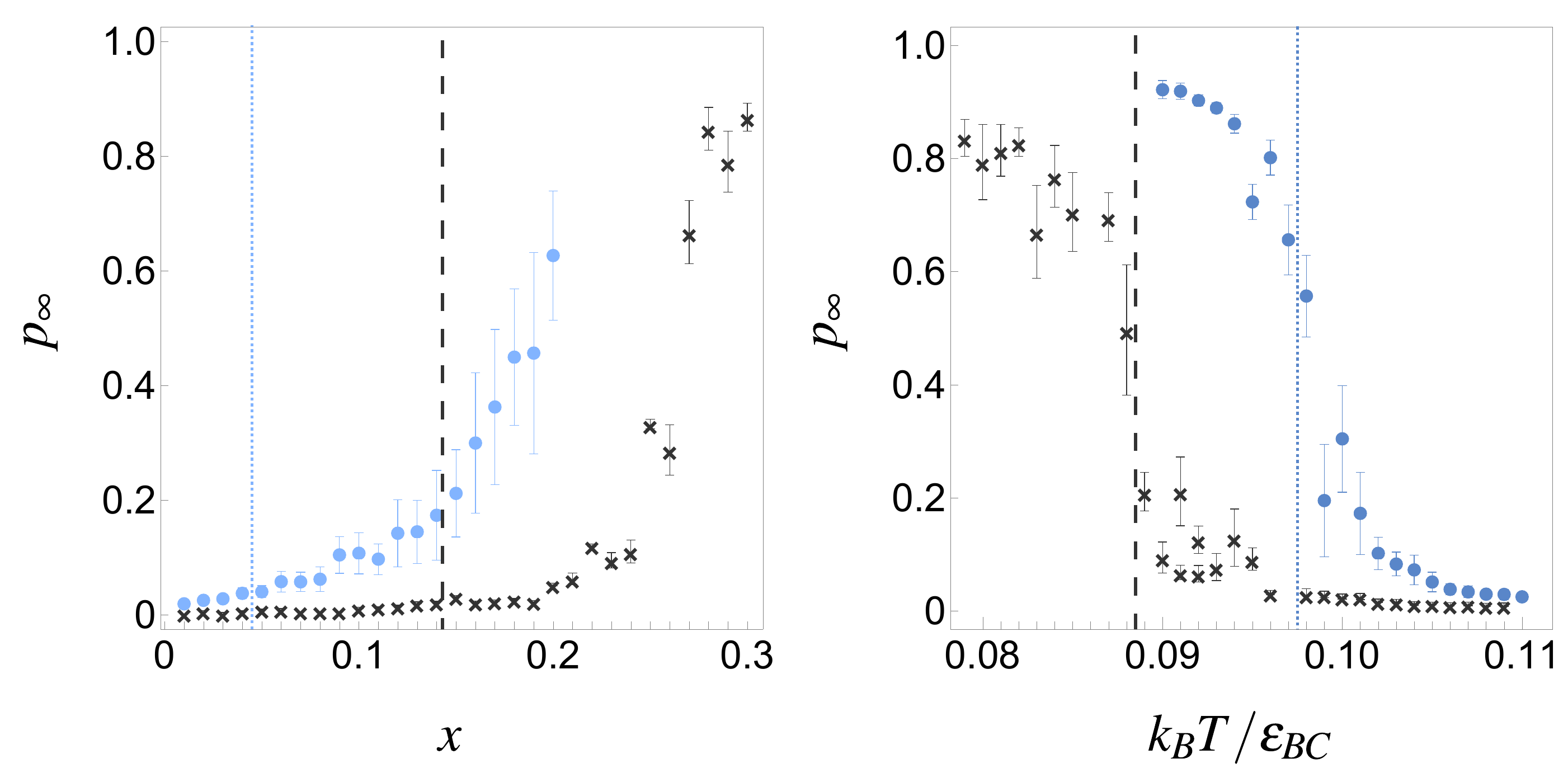}
 \put (1.5,45.5) {a)}
 \put (9.5,0.5) {\protect\includegraphics[width=0.02\textwidth]{images/particles/linkerAABB_plots.png}}
  \put (46,0.5) {\protect\includegraphics[width=0.02\textwidth]{images/particles/coll_plots.png}}
 \put (51,45) {b)}
\end{overpic}
\caption{ 
Fraction of particles and linkers that belong to the largest cluster, $p_\infty$, obtained from simulations for model B. Both plots are for $\rho=0.15$: (a) $k_BT/\epsilon_{BC}=0.08$; $\epsilon_{AA}/\epsilon_{BC}=0.8$ (blue circles) and $\epsilon_{AA}/\epsilon_{BC}=0$ (black crosses); (b) $x=0.3$; $\epsilon_{AA}/\epsilon_{BC}=1$ (blue circles) and $\epsilon_{AA}/\epsilon_{BC}=0$ (black crosses).  The vertical lines signal the composition (in (a)) and the temperature (in (b)) at which the theory predicts percolation to occur.    
}  
\label{fig:simsmodB1}
\end{figure}

Some predictions of the theory for model B were tested (and confirmed) by simulations. Their  results are shown in Fig.~\ref{fig:simsmodB1}. 
The introduction of chains of linkers reduces the fraction of particles at which, at fixed low density and low temperature, percolation occurs, as shown by the comparison between the results for $\epsilon_{AA}/\epsilon_{BC}=0.8$ and for $\epsilon_{AA}/\epsilon_{BC}=0$ in Fig.~\ref{fig:simsmodB1}a). 
The results shown in Fig.~\ref{fig:simsmodB1}b) confirm the increase of the temperature at which percolation occurs (for fixed low density and composition) when chains of linkers are present. 


\subsection{Structural properties at the percolation threshold}

The theory presented in section {\ref{sec:theory}}  can be developed to obtain some properties of the clusters, within its  assumption that clusters are tree like. 
In particular, the percolation threshold equations (\ref{percmodA}) and (\ref{percmodB}) can be expressed in terms of two structural properties: $\langle n_C\rangle$,
the mean number of bonds formed by one particle, that quantifies the number of chains that branch in one particle; and $\langle\ell_{L}\rangle$, the mean length of the chains formed by linkers, that quantifies the extent of chaining.

$\langle n_C \rangle $ is calculated from the number of patches $f_C$ of the particles and the probability $p_C$ that a patch $C$ is bonded:
\begin{equation}
\label{meannc}
\langle n_C \rangle = f_C p_C.
\end{equation}
The chains of linkers that form in both models can be uniquely identified as sequences of consecutive $AA$ bonds. 
 The number of linkers in sequences with $n$ linkers and $n-1$ $AA$ bonds is proportional to $p_{A\to A}^{n-1}$. 
 The  mean length of the chains formed by linkers is then,
\begin{equation}
\label{meanlAA}
\langle \ell_{L} \rangle \equiv \frac{\sum_{n=1}^\infty n p_{A\to A}^{n-1}}{\sum_{n=1}^\infty  p_{A\to A}^{n-1}}=\frac{1}{1-p_{A\to A}}.
\end{equation}
The equations that define the percolation threshold, (\ref{percmodA}) and (\ref{percmodB}) are expressed in terms of $\langle n_c\rangle$ and $\langle\ell_{L}\rangle$, using (\ref{meannc}), (\ref{meanlAA}), and  
(\ref{percmodA}) (for model A) or (\ref{percmodB}) (for model B), to obtain for model A,  
\begin{equation}
\label{scalA}
\langle n_c \rangle^2\langle\ell_{L}\rangle= \frac{2f_C(1-x)}{(f_C-1)x},
\end{equation}
and, for model B,
\begin{equation}
\label{scalB}
\langle n_C \rangle^2\left(2f_B\langle\ell_{L}\rangle-(f_B+1)\right)= \frac{f_Cf_B(1-x)}{(f_C-1)x}.
\end{equation}
This shows that the relation between (the square of) $\langle n_C \rangle$ and $\langle \ell_L\rangle$ at percolation depends only on composition and not on temperature, density or the ratio of energy scales. It is possible to obtain percolated clusters with lower values of $\langle n_C\rangle$ by increasing the length of the chains of linkers. The increase (decrease) of chaining leads to the decrease (increase) of branching. In the original linker-particle model where chains of linkers are absent, it is only possible to control one feature of the onset of  percolation, since $\langle n_C\rangle$ at the percolation threshold is  defined by composition ($\langle n_C\rangle^2= 2f_C(1-x)/((f_C-1)x)$).  The models under study have another degree of control: it is possible to create percolated states with more or less branchng by changing the propensity to chaining (and vice versa). 

\section{Conclusions and discussion}

The present work shows that in linker mediated aggregation of particles, the introduction of self-assembled chains of linkers  gives an extra control over percolation, relatively to the case where chains of linkers are absent. The choice of the specific interactions between linkers (model A or model B), the strength of those interactions (value of $\epsilon_{AA}/\epsilon_{AC}$ or $\epsilon_{AA}/\epsilon_{BC}$) and the bonding volumes, can be used to tune the composition, the temperature and the density at which percolation occurs, and to change the structure of the percolated clusters. 

In model B, where chaining of linkers and branching through bonds between chains and particles are independent, it is  possible  to reach percolation at both low and high concentrations of particles: a vanishing amount of particles (linkers)  is enough for percolation to occur in systems with a large fraction of linkers (particles). 
 Model A - where competition between chaining of linkers and branching of chains (through their bonds to particles) is present - exhibits a more complex behavior, that depends on the ground state of the model. For $\epsilon_{AA}/\epsilon_{AC}<2$, branching is energetically  favoured, and percolation is extended to low particle composition. The temperature at which, at low compositions, percolation disappears depends non monotonically on $\epsilon_{AA}/\epsilon_{AC}$, being maximum for $\epsilon_{AA} \approx \epsilon_{AC}$. This non-monotonic behavior is related to the energy costs of disrupting the percolated network, either  by breaking  the chains or by breaking the bonds between chains and particles. Percolation occurs in model A with $\epsilon_{AA}>2\epsilon_{AC}$ only when the entropic gain of branching is increased. In this case, percolation by heating and closed loops of percolated phases in  temperature-composition diagrams are obtained.
 The number of branching bonds per particle is related to  the mean length of chains of linkers  through a function that depends only on composition. This means that   composition controls the structure of the percolated clusters:  it is possible to obtain clusters with larger (smaller) chains at the cost of decreasing (increasing) branching. 
 The simulations performed for model A (for $\epsilon_{AA}<2\epsilon_{AC}$) and for model B have confirmed the qualitative differences predicted by the theory for the percolation diagrams obtained when chains of linkers are present.
 
 Percolation is only a necessary condition to obtain an equilibrium gel. The phase diagrams of the models of this work are needed to know the thermodynamic conditions at which percolation happens in single phase regions. These calculations are left for future work, but some hints can be found in previous studies \cite{teixeira2021,delasHeras2011a,delasHeras2011b}. 
    The case of model A (or model B with $f_B=2$) with $\epsilon_{AA}=0$ (particle-linker mixture with no chains) and $f_C=3$ was investigated in \cite{teixeira2021}. Percolated liquids, not affected by phase separation, were found at intermediate compositions, low densities and low temperatures (see Figs.~4, 5 and 6 in \cite{teixeira2021}).  Moreover, in \cite{delasHeras2011b} the phase diagram of model A with $\epsilon_{AA}=\epsilon_{AC}$ and $f_C=3$ shows that at high linker concentrations the percolated liquids have less tendency to phase separate with a vapour (see Fig.~6 in \cite{delasHeras2011b}). Even in a mixture of patchy particles with 2 and 4 patches, all equal and all forming bonds, large regions of a single phase percolated liquid are found \cite{delasHeras2011a}. Therefore, it is much likely that the models studied in this work will exhibit regions of the phase diagrams with a single percolated liquid phase.

Finally, it should be mentioned that, to complete the study of equilibrium gels, the onset of mechanical stability in the percolated network has to be addressed \cite{Jacobs1995,Tsurusawa2019}. The emergence of rigidity in equilibrium self-assembling thermodynamic systems is poorly understood \cite{Zhang2019}. However, recent developments of a theory that describes the
mechanical properties of ideal reversible polymer networks at thermodynamic equilibrium, may apply to the linker-particle systems
\cite{parada2018}
, if properly generalized to the case of a mixture.

\begin{acknowledgments}
We acknowledge financial support from the Portuguese Foundation for Science and Technology (FCT) under Contracts no. PTDC/FIS-MAC/28146/2017 (LISBOA-01-0145-FEDER-028146), PTDC/FIS-MAC/5689/2020, CEECIND/00586/2017, UIDB/00618/2020,  and UIDP/00618/2020.
\end{acknowledgments}

\appendix
\begin{figure}[htb]
\begin{overpic}[width=0.35\textwidth]{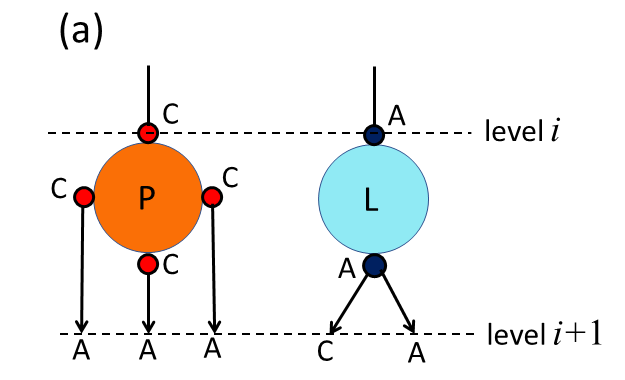}
\end{overpic}
\begin{overpic}[width=0.5\textwidth]{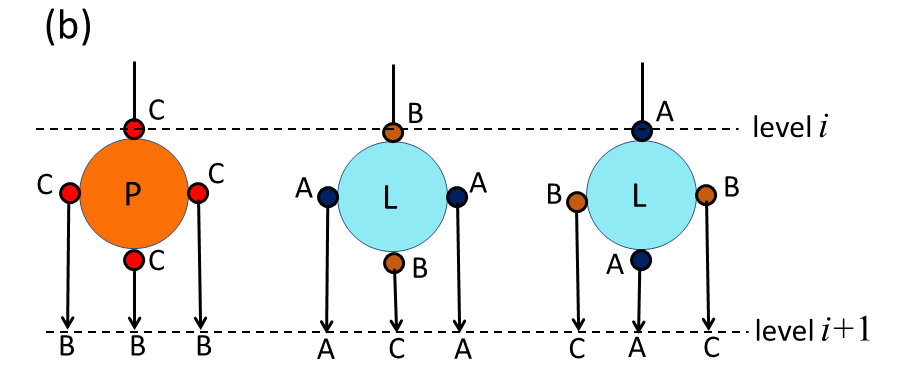}
\end{overpic}
\caption{Schematic representation of the levels of a tree like cluster formed by particles and linkers that is the basis to obtain the expressions of matrices $\tilde T$ (\ref{matrixA},\ref{matrixB}). a) Model A; b) Model B.}
\label{fig:matrixT}
\end{figure}
\section{\label{appMatrix}Matrix $\tilde T$}
Let us consider the levels of a tree like cluster (with no loops) formed by linkers and particles connected through the bonding of  its patches, as described in section \ref{sec:theory}. $b_{i,\gamma}$ is the number of patches of type $\gamma$ that belong to particles or linkers in level $i$ and  that are bonded to the previous level. Matrix
 $\tilde T$ in (\ref{matrixA}) and (\ref{matrixB}) is obtained from the relation between the number of patches $b_{i,\gamma}$ of different levels. 
 Figure \ref{fig:matrixT} represents, for models $A$ and $B$, all the possible ways of obtaining a patch of level $i+1$ that is bonded to level $i$, from the patches of level $i$ that are bonded to level $i-1$.
\subsection{Model A}
In model A only bonds $AA$ and $AC$ are allowed. A patch $C$ at level $i+1$ can only be obtained, with probability $p_{A\to C}$, from a bond that originates in a patch $A$ of a linker of level $i$; this linker is connected to level $i-1$ through its other patch $A$. Therefore, 
\begin{equation}
\label{bcmodelA}
    b_{i+1,C}=p_{A\to C} b_{i,A}.
\end{equation}
A patch A in level $i+1$ can be obtained in two ways: (i) with probability $p_{A\to A}$, from a bond that originates in a patch A of a linker of level $i$; this linker is connected to level $i-1$ through its other patch A; (ii) with probability $p_{C\to A}$, from a bond that originates in one of the available $f_C-1$ patches  of a particle of level $i$; this particle is connected to level $i-1$  by its remaining C patch. As a consequence,
\begin{equation}
\label{bamodelA}
    b_{i+1,A}=p_{A\to A} b_{i,A}+(f_C-1)p_{C\to A} b_{i,C}.
\end{equation}
The elements of matrix $\tilde T$ (\ref{matrixA}) are the coefficients of the generalized geometric progression expressed in (\ref{bcmodelA},\ref{bamodelA}).

\subsection{Model B}
In model B, bonds $AA$ and $BC$ are allowed and 3 different type of patches are involved in bonds.

A patch B in level $i+1$ can be obtained,  with probability $p_{C\to B}$, from a bond that originates in one of the available $f_C-1$ patches  of a particle of level $i$; this particle is connected to level $i-1$  by its remaining C patch. As a consequence,
\begin{equation}
\label{bbmodelB}
    b_{i+1,B}=(f_C-1) p_{C\to B} b_{i,C}.
\end{equation}
A patch A in level $i+1$ can be obtained, with probability $p_{A\to A}$, from a bond that originates in a patch A of a linker of level $i$; this linker can be bonded to level $i-1$  by  a patch $A$ or by a patch $B$ (in which case, the 2 patches $A$ of the linker of level $i$ are available to bond to the patch $A$ of level $i+1$).  As a consequence,
\begin{equation}
\label{bamodelB}
    b_{i+1,A}=p_{A\to A} b_{i,A}+2p_{A\to A} b_{i,B}.
\end{equation}
Finally, a patch C in level $i+1$ can be obtained, with probability $p_{B\to C}$, from a bond that originates in a patch B of a linker of level $i$; this linker can be bonded to level $i-1$ by: (i) a patch $A$, in which case the $f_B$ patches B of the linker of level $i$ are available to bond to the patch C of level $i+1$; (ii)  by a patch $B$, in which case  only $f_B-1$ patches of the linker of level $i$ are available to bond to the patch $C$ of level $i+1$.
As a consequence,
\begin{equation}
\label{bcmodelB}
    b_{i+1,C}=f_B p_{B\to C} b_{i,A}+(f_B-1)p_{B\to C} b_{i,B}.
\end{equation}

The elements of matrix $\tilde T$ (\ref{matrixB}) are the coefficients of the generalized geometric progression expressed in (\ref{bbmodelB},\ref{bamodelB},\ref{bcmodelB}).

\section{\label{apppercB}Percolation threshold of model B}

The percolation threshold is obtained  when the  maximum absolute value of all eigenvalues of matrix $\tilde T$ equals 1.
Here we show that the percolation threshold  of model B can be found by simply searching the conditions for which  $\lambda=1$ is an eigenvalue of (\ref{matrixB}), as was done in (\ref{percmodB}) and in subsequent calculations.  
The eigenvalues $\lambda$ of (\ref{matrixB}) are the solutions of,
\begin{equation}
\label{percmodB3}
\lambda^3-p_A\lambda^2-(f_B-1)z\lambda-(f_B+1)p_Az=0,
\end{equation}
where $z=(f_C-1) p_{B\to C} p_{C\to B}$ and $p_A=p_{A\to A}$. 
\begin{itemize}

\item[1.] {We will first show that if $\lambda=1$ is a solution of (\ref{percmodB3}), this solution is the one with the largest absolute value.

If $\lambda=1$ is a solution of (\ref{percmodB3}), then  (\ref{percmodB3}) can be rewritten as,
\begin{equation}
\label{factorize}
(\lambda-1)(\lambda^2+b\lambda +c)=0,
\end{equation}
where $b=1-p_A$, $b-c=(f_B-1)z$ and $c=(f_B+1)p_Az$. From these equalities, $c$ can be expressed as a function of $p_A$,
\begin{equation}
\label{cpA}
c=\frac{(f_B+1)(1-p_A)p_A}{f_B-1+(f_B+1)p_A}.
\end{equation}
The other eigenvalues, $\lambda_\pm$, are,
\begin{equation}
\label{otherev}
\lambda_\pm=\frac{-b\pm\sqrt{b^2-4c}}{2}.
\end{equation}
If these eigenvalues are not real  (i.e. if $b^2-4c<0$), then $\lambda=1$ is the only real solution of (\ref{percmodB3}) and therefore is the one  with the largest absolute value.
If they are real, one can use the relation between $c$, $b$ and $p_A$ and,  taking into account that $0<p_A<1$ , obtain their absolute values, 
\begin{equation}
\label{lambdap}
|\lambda_+|= \frac{1-p_A}{2}\left[ 1- \sqrt{1-G(f_B,p_A)}\right],
\end{equation}
\begin{equation}
\label{lambdam}
|\lambda_-|= \frac{1-p_A}{2}\left[ 1+ \sqrt{1-G(f_B,p_A)}\right],
\end{equation}
where,
\begin{equation}
G(f_B,p_A)=\frac{4(f_B+1)p_A}{(1-p_A)\left[f_B-1+(f_B+1)p_A\right]}.
\end{equation}
Since $\frac{1-p_A}{2}<1/2$, $\left[ 1- \sqrt{1-G(f_B,p_A)}\right] <1$ and $\left[ 1+ \sqrt{1-G(f_B,p_A)}\right]<2$, we conclude that $|\lambda_+|<1$ and $|\lambda_-|<1$. Therefore, if there is an eigenvalue $\lambda=1$, the other two, if they exist, have absolute values lower than 1.
}

\item[2.]{We will show that if $\lambda=-1$ is a solution of (\ref{percmodB3}) then there is another eigenvalue whose absolute value is larger than 1, and therefore $\lambda=-1$ does not define a percolation threshold. 

Assume  that one of the solutions of (\ref{percmodB3}) is $\lambda=-1$, so that  (\ref{percmodB3}) can then be rewritten as,
\begin{equation}
\label{factorize2}
(\lambda+1)(\lambda^2+b\lambda +c)=0,
\end{equation}
where $b=-1-p_A$, $b+c=-(f_B-1)z$ and $c=-(f_B+1)p_Az$. From these equalities, $c$ and $z$ can be expressed as a function of $p_A$,
\begin{equation}
\label{cpB}
c=-\frac{(f_B+1)(1+p_A)p_A}{f_B-1-(f_B+1)p_A}.
\end{equation}
\begin{equation}
\label{zpA}
z=\frac{1+p_A}{f_B-1-(f_B+1)p_A}.
\end{equation}
Since $z>0$, (\ref{zpA}) is only valid if $p_A<\frac{f_B-1}{f_B+1}$, which means that $\lambda=-1$ can be a solution of (\ref{percmodB3}) only when $p_A$ satisfies this inequality. 
The other eigenvalues, $\lambda_\pm$, are,
\begin{equation}
\label{otherev2}
\lambda_\pm=\frac{-b\pm\sqrt{b^2-4c}}{2}=(1+p_A) \frac{1\pm\sqrt{1+F(p_A,f_B)}}{2},
\end{equation}
where
\begin{equation}
F(f_B,p_A)=\frac{4(f_B+1)p_A}{(1+p_A)\left[f_B-1-(f_B+1)p_A\right]}.
\end{equation}
For $p_A<\frac{f_B-1}{f_B+1}$ (i.e. when $\lambda=-1$ can be one of the solutions of (\ref{percmodB3})), $F(f_B,p_B)>0$ and so $\lambda_\pm$ are always real. 
The absolute value of $\lambda_+$ is,
\begin{equation}
\label{eigv2}
|\lambda_+|=(1+p_A) \frac{1+\sqrt{1+F(p_A,f_B)}}{2}
\end{equation}
Since $1+p_A>1$ and $1+\sqrt{1+F(p_A,f_B)}>2$, then $|\lambda_+|>1$. As a consequence $\lambda=-1$ is never the eigenvalue of (\ref{matrixB}) with maximum absolute value and cannot be used to obtain the percolation threshold.
} 
\end{itemize}
We conclude that the general definition for the percolation threshold, that is, the thermodynamic conditions for which the maximum absolute value of all eigenvalue of matrix $\tilde T$ is 1, reduces, for model B, to the conditions at which one of the eigenvalues is equal to 1.


\end{document}